# Expanding the Timeline for Earth's Photosynthetic Red Edge Biosignature

Jack T. O'Malley-James[*,1] and Lisa Kaltenegger[1,2]
[1]Carl Sagan Institute at Cornell University, [2]Astronomy department, Ithaca, NY 14853, USA
jomalleyjames@astro.cornell.edu
*Corresponding author

ABSTRACT
When Carl Sagan observed the Earth during a Gallileo fly-by in 1993, he found a widely distributed surface pigment with a sharp reflection edge in the red part of the spectrum, which, together with the abundance of gaseous oxygen and methane in extreme thermodynamic disequilibrium, were strongly suggestive of the presence of life on Earth. This widespread pigmentation that could not be explained by geological processes alone, is caused by the cellular structure of vegetation – a mechanism for potentially limiting damage to chlorophyll and/or limiting water loss. The distinctive increase in the red portion of Earth's global reflectance spectrum is called the vegetation red edge in astrobiology literature and is one of the proposed surface biosignatures to search for on exoplanets and exomoons. Earth's surface vegetation has only been widespread for about half a billion years, providing a surface biosignature for approximately $1/9^{th}$ our planet's lifetime. However, as chlorophyll is present in many forms of life on Earth, like cyanobacteria, algae, lichen, corals, as well as leafy vegetation, such a spectral red edge feature could indicate a wide range of life, expanding its use for the search for surface biosignatures beyond vegetation alone to a time long before vegetation became widespread on Earth. We show how lichens could extend the presence of Earth's red edge surface biofeature to 1.2 Gyr ago, while ocean surface algae and cyanobacteria could extend it to over 2 Gyr ago, expanding the use of a photosynthetic red edge to earlier times in Earth's history.

## 1. INTRODUCTION

Among several thousand detected exoplanets that now provide a first glimpse of the diversity of other worlds (e.g., reviewed in Winn & Fabrycky 2015) are the first small exoplanets that could potentially be habitable (see, e.g., Batalha 2014; Kane et al. 2016). Signs of life that modify the atmosphere or the surface of a planet, and thus can be remotely detectable, are key to finding life on exoplanets or exomoons (see e.g. review by Kaltenegger 2017). Remote direct detection of surface life in reflected light from exoplanets becomes possible when organisms modify the detectable reflectivity of the surface (e.g. by influencing surface colors). Land vegetation is commonly cited as such a surface biosignature, indicative of life (e.g. Sagan et al. 1993).

The most abundant surface reflectance feature indicating the presence of life on present-day Earth is land vegetation. Vegetation exhibits a strong increase in reflectance at ~700 nm; a feature commonly called the vegetation red edge (VRE) in exoplanet biosignature studies (see e.g. Seager et al. 2005, Arnold et al. 2009; Fujii et al. 2010; Schwieterman et al. 2018). The strength of this increase varies between plant species (e.g. O'Malley-James & Kaltenegger 2018), but is typically ~50% for present-day common deciduous vegetation. Vegetation covers about 60% of present-day Earth's land surface, thus the VRE can be seen in Earth's globally averaged reflectance spectrum as an increase of a few percent (e.g. Seager et al. 2005; Montañés-Rodríguez et al. 2006, Arnold et al. 2009; O'Malley-James & Kaltenegger 2018). Note that a VRE surface biosignature similar to that on the present-day Earth would be difficult to detect without very high-precision instruments (e.g. Arnold et al. 2009). The vegetation red edge has been proposed as a remotely detectable surface biosignature for habitable planets, which will be challenging but very interesting to detect over interstellar distances (see e.g. Sagan et al. 1993, Seager et al. 2005, O'Malley-James & Kaltenegger 2018).

Vegetation is defined as assemblages of plant life, where plants are defined as photosynthetic eukaryotes, consisting of the flowering plants, conifers and other gymnosperms, ferns, clubmosses, hornworts, liverworts, mosses and green algae. The exact wavelength and strength of the spectroscopic VRE depends on the plant species and environment (see e.g. Kiang et al. 2007 and discussion in Rothschild 2008). All of these organisms have been responsible for Earth's vegetation reflectance feature through geological time (O'Malley-James & Kaltenegger 2018) since land plants first appeared ~725 – 500 million years ago (Zimmer et al. 2007; Magallón et al.

2013).

Yet, even on Earth, chlorophyll-containing photosynthetic structures that exhibit a red edge feature are present in organisms other than vegetation (see Fig. 1); including lichens (e.g. Clark 2007), corals (e.g. Roelfsema & Phinn 2006), algae and cyanobacteria (Hegde et al. 2015), which developed long before vegetation became widespread. Widening the commonly held conception of the vegetation red edge to encompass these others forms of photosynthetic life – a photosynthetic red edge – extends the timeline of possible detection of photosynthesic life on Earth to up to 2 Gyr ago, as well as expanding the applicability of this biosignature to a wider range of exoplanets, including rocky ocean planets with large water surfaces.

Chlorophyll-a and -b, the primary molecules that absorb energy from light and use it to drive oxygenic photosynthesis (converting $H_2O$ and $CO_2$ into sugars and $O_2$), have a distinct reflectance profile: Chlorophyll *a* has its strongest absorption peaks at 0.450 μm, chlorophyll *b* has its main absorption peak at 0.680 μm (e.g. Gates et al. 1965; Kiang et al. 2007, Seager et al. 2005 and references therein). A thorough analysis of the likelihood of oxygenic photosynthesis arising elsewhere is given by Wolstencroft & Raven (2002) and Rothschild (2008).

Here, we explore whether a photosynthetic red edge feature in an (exo)planet's spectrum could be produced by organisms aside from vegetation, whether such features could be detectable – expanding the range of organisms that could be uncovered by detecting a red edge – and whether we could distinguish the organisms producing the red edge from vegetation when observing an exoplanet. This means we model disk integrated observations with high, as well as limited, spectral resolution. We use the reflectance spectra of four different photosynthetic organisms, which could have provided widespread surface coverage on our Earth as template for Earth-like planets: We use cyanobacteria, algae, lichen, as well as deciduous vegetation to assess how the strength of a planet's photosynthetic red edge signature would change if different types of organisms were dominant.

Cyanobacteria have populated Earth for at least 2 billion years, with some evidence suggesting they could be as old as ~3.5 billion years (Knoll 2008; Brasier et al. 2015), while the earliest fossils attributed to algae are around 1 billion years old (Knoll 2008). Lichen may also have emerged about 1 billion years ago (Horodyski & Knauth, 1994; Raven 1997; Knauth & Kennedy 2009), while corals and modern vegetation only appeared about ~725 – 500 million years ago (Pratt et al. 2001).

We model how Earth's (or an Earth-like planet's) spectrum would change as a result of a change in dominant organism. A reduction in the covered surface as well as an increase in cloud coverage both reduce the strength of any surface feature, while an increase in covered surface area (e.g. a bigger Super-Earth) and reduction of cloud coverage (e.g. a drier planet) will increase it. We show disk-integrated spectra similar to the observation geometry obtained shortly before or after secondary eclipse or during direct imaging of the planet itself, for clear sky as well as Earth-like cloud coverage (see Figure 2).

2. METHODS

We model planetary spectra with different surface coverage for organisms using chlorophyll with a present-day Earth atmosphere using *EXO-Prime* (Kaltenegger & Sasselov 2009); a coupled 1D radiative-convective atmosphere code developed for rocky exoplanets, which models an Earth-like exoplanet's atmosphere, its spectrum (see e.g. Kaltenegger 2010; Rugheimer et al. 2013; 2015, 2017) as well as the UV environment on its surface (see Rugheimer et al. 2015; O'Malley-James & Kaltenegger 2017; Kozakis & Kaltenegger 2019).

Our radiative transfer model is based on a model that was originally developed to model the Earth's atmospheric spectra (Traub & Stier 1976) and has since been used extensively for analyzing high-resolution Fourier transform spectra from ongoing stratospheric balloon-based observations to study the photochemistry and transport of the Earth's atmosphere (for example, Jucks et al. 1998). Our line-by-line radiative transfer code has also been used for numerous full planetary disk modeling studies, both for theoretical studies (e.g., Des Marais et al. 2002; Kaltenegger et al. 2010; Rugheimer et al. 2015, 2017) and fitting observed earthshine spectra (e.g., Woolf et al. 2002; Turnbull et al. 2006; Kaltenegger et al. 2007; Rugheimer et al. 2013). We divide the atmosphere into 60 thin layers from 0 to 100 km in altitude. The spectrum is calculated at very high spectral resolution, with several points per line width, where the line shapes and widths are computed using Doppler and pressure broadening on a line-by-line basis for each layer in the model atmosphere.

We use a simple geometrical model in which the spherical Earth is modeled with a plane-parallel atmosphere and a single angle of incidence and reflection (visible) or emission (thermal infrared). This angle is selected to give the best analytical approximation to the integrated-Earth air mass factor of 2 for a nominal illumination (quadrature); the zenith angle of this ray is 60 degrees.

The overall high-resolution spectrum is calculated at a resolution of 0.1 wavenumbers and smeared to lower resolution. For reference and further explanation concerning the code, the reader is referred to our calculation of a complete set of molecular constituent spectra, for a wide range of mixing ratios, for the present-day Earth pressure-temperature profile, and for the visible to thermal infrared, in Des Marais et al. (2002) and Kaltenegger et al. (2007).

To explore how a range of organisms using chlorophyll affect the disk integrated spectrum of a planet and specifically the photosynthetic red edge signal, we use normalized reflectance spectra (Roelfsema & Phinn 2006; Clark 2007; Cartaxana et al. 2017) for the range of photosynthetic organisms shown in Figure 1. For aquatic organisms (algae, cyanobacteria) we assume these are living on the ocean surface; hence, water attenuation and transmission can be neglected for these cases.

We will shortly discuss aquatic organisms that are not surface organisms in the Discussion section. We did not include them in our full study because, while much is known about Earth's oceans, in an exoplanet context it could potentially be misleading to assume absorption and transmission properties using an Earth-analog ocean, especially if the ocean's composition and the host star differ from our own planet's.

We first model a scenario where one organism covers the entire surface of an Earth-size planet with a present-day Earth composition atmosphere, modeling both clear-sky and 50% cloud-fraction scenarios analogous to the present day Earth. Note that while caution should be applied to this initial hypothesis of a planet completely covered by one single phototrophic organism – on Earth today, several different species coexist on the planet, making the surface coverage of the dominant species less than 100% – it is a scenario that also should not be ruled out. We use this scenario, 100% surface coverage, as a starting point in our simulations to explore whether the effect is detectable in principle. We then refine this initial model to represent an Earth analog, by combining a mixture of surface types, which includes 70% of the planetary surface as uninhabited ocean. The remaining surface is composed of 2% coast and 28% land. The land surface consists of 60% vegetation, 9% granite, 9% basalt, 15% snow, and 7% sand, which reproduces Earth's disk integrated spectrum (following Kaltenegger et al. 2007).

Note that while a snapshot of Earth's reflectance spectrum can vary considerably at any given moment of time as the planet rotates and different surface compositions come into view (see e.g. Ford et al. 2001, Palle et al 2008), exoplanet observations will need to be collected over tens of hours to several days to allow the level of detailed study required to identify surface biosignatures. Therefore, a disk integrated model is similar to the quality of such future data. While observations of a small number of sufficiently photon-rich exoplanet targets with large telescopes may provide detailed spectroscopic data that could capture changes on an exoplanet's observable surface as it rotates, in general observing a planet as one dot of light with integration times of tens of hours to days would provide measurements that are similar to a disk integrated spectra combining the different areas of a planet (see e.g. Kaltenegger et al. (2007) and Rugheimer (2013) for validation of Earth's disk integrated spectra with remote Earth observations). We then substitute modern vegetation with the selected organism to investigate the impact of different chlorophyll-containing life on a planet's spectrum. While we have no guidelines for whether another habitable exoplanet would have a similar land-ocean distribution to the present-day Earth, we use this scenario to show the influence of surface coverage on the signal representative of a wide model grid in Table 1.

How changes in climate could alter a planet's cloud fraction is still debated: For example, as discussed for Earth's evolution in the literature, warmer climates could increase humidity, favoring increased cloud formation (see e.g. Sellwood et al. 2000), but higher temperatures could also reduce nutrient cycling in the oceans, reducing the rate of biologically-produced cloud condensation nuclei, leading to optically thinner, shorter-lived clouds (Kump & Pollard 2008). Even if the overall cloud fraction remained approximately constant, cloud distributions could change (see e.g. Brierley et al. 2009, which shows how in a warmer climate high-cloud fractions could increase and low-cloud fractions could decrease). While the discussion is on-going, we model a 50% cloud coverage, based on present-day Earth's cloud coverage, modelled at the height of Earth-clouds with similar fractions, such that the model is consistent with Earthshine data (see Kaltenegger et al. 2007).

### 3. RESULTS & DISCUSSION

*3.1 The Photosynthetic Red edge feature can indicate*

*a wide range of organisms*

Organisms, which contain chlorophyll, besides vegetation, could also produce similar red edge surface biosignatures for similar surface coverage (see Fig. 2 and Table 1), depending on how widespread their surface coverage is. We do not have data for the surface coverage of e.g. lichen or cyanobacteria for a young Earth. Table 1 shows that for similar surface coverage the Photosynthetic Red Edge signal of other organisms that could be dominant on the surface of an exoplanet can be similar in strength to the signal produced by modern vegetation for Earth in our models, which is approximated using deciduous tree reflectance producing an estimated reflectance increase of ~4% (Table 1), falling within the lower end of the range of values (1-10%) given for Earth's vegetation red edge (see e.g. review by Arnold 2009). Figure 1 shows that individually the different organisms can be distinguished with high spectral resolution. However, once we add a present-day Earth atmosphere as well as clouds to the model (Figure 2), the individually distinguishing slope of the reflectivity of the organisms is no longer apparent. Thus a red edge detection, while not being specific to any one form of photosynthetic organism, can indicate a wider range of organisms than only vegetation.

*3.2 Lichens could have produced a Photosynthetic Red Edge surface biosignature before vegetation emerged on a young Earth*

Lichens may have colonized the land between 0.85 and 1.2 Gyr ago (Horodyski & Knauth, 1994; Raven 1997; Knauth & Kennedy 2009) causing a "non-vegetation" red edge signature before Earth gained its vegetation red edge about 0.5 Gyr ago. As shown in Fig. 1, lichens have red edge strengths of ~20%. This is only about half the strength of that of modern vegetation, resulting in a weaker globally averaged red edge signature on a lichen-dominated planet (see Fig. 2, Tab. 1), compared to present-day vegetation for similar surface coverage. Note that the vegetation red edge was also weaker initially and has changed in strength over the past 500 Myr and would likely have had a similar strength to lichens when the first land plants emerged; see O'Malley-James & Kaltenegger (2018) for details. However lichens could have provided an observable red edge for a younger Earth before vegetation became widespread on land, extending the time the red edge surface feature was detectable for our own planet to about a billion years ago.

*3.3 A widespread surface algae and cyanobacteria biosphere could have produced a Photosynthetic Red Edge surface biosignature before vegetation on a young Earth*

Cyanobacteria may have been widespread between ~3.5 and 2 Gyr ago (Knoll 2008; Brasier et al. 2015) causing a photosynthetic red edge signature before Earth surface vegetation became wide-spread about 0.5 Gyr ago. As shown in Fig. 1, cyanobacteria have red edge strengths of ~25%. This is only about half the strength of that of modern vegetation, resulting in a weaker globally averaged red edge signature on a planet with ocean surface dwelling algae and cyanobacteria (see Fig. 2, Tab. 1), compared to present-day vegetation for similar surface coverage.

Note that we assume surface dwelling algae and cyanobacteria in our models. In a water body, microbial photosynthesizers can live at any depth within the photic zone (defined as the depth to which sufficient light penetrates to drive photosynthesis), given adequate temperatures and nutrient availability; however, the highest concentrations of such life tend to be found at or near the surface in ideal growing conditions. Note that the depth of the photic zone would vary based on star type and the orbital distance of the planet. Hence, microscopic ocean surface-dwelling organisms could have provided an observable red edge for a younger Earth before vegetation or lichens became widespread on land, extending the time the red edge surface feature was detectable for our own planet to at least 2 billion years ago.

*3.4. Other life forms can also generate a Photosynthetic Red Edge feature.*

Other forms of life can also show a red edge, such as some types of corals, which show up to ~65% increase in reflectance in the visible (Clark 2007). For subsurface aquatic photosynthetic life, water attenuation and transmission would need to be accounted for, which would reduce the detectable reflectivity feature with increasing amounts of overlying water and depends on the ocean's composition, suspended particles as well as the host star. While much is known about Earth's oceans, in an exoplanet context ocean composition is unknown and thus have not been included in our detailed study.

On Earth, corals do not occupy a significant enough surface fraction to produce a strong global red edge feature on Earth. However, this does not exclude the possibility of such forms of life being abundant on another habitable planet. For example, for a similar surface coverage, a coral-like organism has the potential to produce a stronger red edge feature than modern land vegetation (see Fig. 1), although this would depend on the water depth, which would reduce the strength of subsurface reflection features.

For completeness note that even photosynthetic animals like the sea slug *Elysia viridis* (e.g. Cartaxana et al. 2017) show an red edge increase of ~35% (see Figure 1). These features are due to symbiotic relationships with green algae and/or the direct incorporation of chlorophyll into tissues but are not expected to provide a widespread surface coverage.

## 4. CONCLUSIONS

Using Earth, with its diverse biota, as our Rosetta Stone to identify life on other worlds, we show that the Photosynthetic Red Edge (PRE) can indicate a wide range of organisms on a habitable exoplanet, in addition to surface vegetation. While oxygenic photosynthesis, based on chlorophyll-like structures, may be favored by evolution on other worlds (Wolstencroft & Raven (2002); Rothschild (2008)), whether or not vegetation becomes the dominant surface coverage on exoplanets or exomoons is unknown.

We show and compare disk-integrated spectra of Earth-like planets assuming that they are dominated by four different organisms, which use chlorophyll: cyanobacteria, algae, lichen, as well as deciduous vegetation, expanding the timeline for the red edge surface biosignature from 500 Million years, based on widespread surface vegetation, back in Earth's history to at least 2 billion years ago due to other biota which use photosynthesis.

When adding present-day Earth's atmosphere as well as clouds to the model (Figure 2), the individually distinguishing slope of the reflectivity of the different organisms (Fig. 1) is no longer apparent in a low resolution reflection spectrum. Thus a photosynthetic red edge surface biosignature detection can indicate a wider range of organisms than just vegetation.

A PRE biosignature similar to that on the present-day Earth would be difficult to detect without very high-precision instruments on exoplanets. However, our models show that the PRE signature can increase for different organisms that use chlorophyll, as well as with increasing surface fraction of an organism and decreasing cloud coverage on a planet.

Depending on the unknown surface coverage, organisms like lichens and algae at the ocean surface could have provided a detectable surface PRE feature for a younger Earth about 1 billion years ago, and cyanobacteria up to at least 2 billion years ago, long before land vegetation became widespread about 750 to 500 Million years ago, which dominates present-day Earth's PRE signal. This expands the use of a photosynthetic red edge surface biofeature to earlier times in Earth's history as well as to a wider range of habitable extrasolar planet scenarios.

**Acknowledgements** The authors would like to thank the anonymous reviewer for constructive comments that helped to improve the manuscript. The authors acknowledge the *USGS Digital Spectral Library, ASTER Spectral Library* and the *Joint Fire Science Program and the Color catalog of Life at the Carl Sagan Institute Cornell* spectral library. We also acknowledge funding from the Simons Foundation (290357, Kaltenegger).

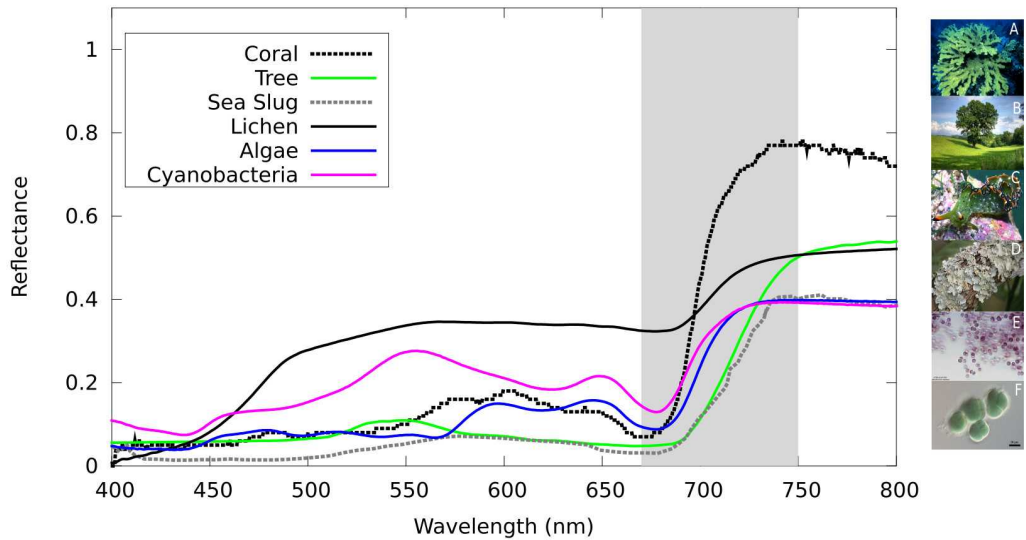

**Figure 1**. Examples of red edge features – the increase in reflectance caused by chlorophyll, highlighted in the shaded region – exhibited by (**A**) corals, (**B**) deciduous vegetation (trees; representative of the present-day red edge feature in Earth's spectrum), (**C**) the photosynthetic sea slug, *Elysia viridis,* (**D**) lichen (*Acarospora sp.*), (**E**) algae (*Rhodosorus marinus*), (**F**) cyanobacteria (*Chroococcidiopsis sp.*).

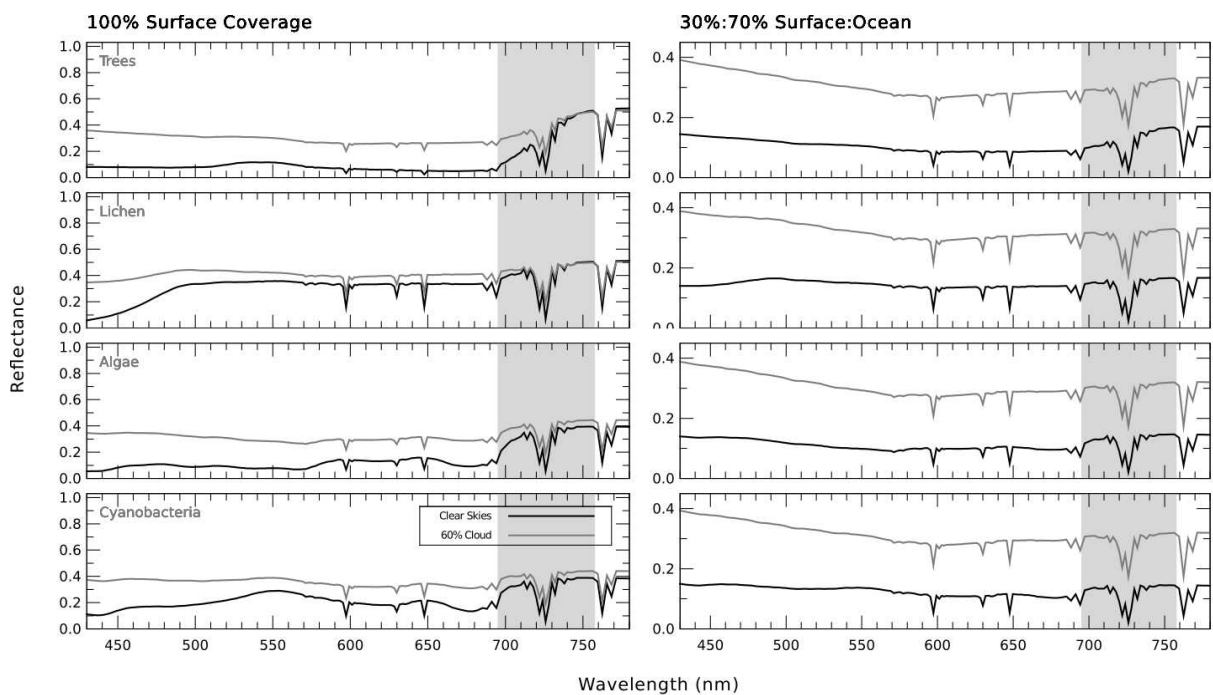

**Figure 2**. Model spectra showing how an Earth-like atmosphere influences the red edge strengths for the examples from Fig. 1 (excluding the sea slug, *Elysia viridis* and corals), for a clear atmosphere (black line) and an Earth-like cloud fraction (grey line). (*Left*) assuming 100% of the planet's surface is covered by the organism; (right) similar surface coverage as present-day Earth, replacing vegetation with the selected organism. Note that algae and cyanobacteria are assumed to be at the ocean surface.

| *Species* | 100% Surface Cover [clear] (%) | 100% Surface cover [50% cloud] (%) | Earth analog surface [clear] (%) | Earth analog surface [50% cloud] (%) |
|---|---|---|---|---|
| *Trees* | 40 | 20 | 9 | 4 |
| *Lichen* | 18 | 9 | 2 | 1 |
| *Algae* | 23 | 11 | 6 | 3 |
| *Cyanobacteria* | 20 | 10 | 4 | 2 |

**Table 1.** The percentage increase in reflected flux at the red edge. Results show the increase in visible flux at ~700 nm for the globally averaged model planet spectra plotted in Figure 2 for: (left) 100% of the planet's surface is covered by the organism; (right) similar surface coverage to the present-day Earth, replacing vegetation with the selected organism.